\documentclass[modern]{aastex61}
\usepackage{natbib}
\usepackage{placeins}
\usepackage{hyperref}
\usepackage{rotfloat}
\usepackage{multirow}
\usepackage{figs}

\newcommand{\SOFIA}{{\it SOFIA}}

\makeatletter
\def \blfootnote{\xdef\@thefnmark{}\@footnotetext}
\makeatother

\begin{document}
\title{Searching for Cool Dust: II. Infrared Imaging of\\
  The OH/IR Supergiants, NML Cyg, VX Sgr, S Per\\
  and the Normal Red Supergiants RS Per and T~Per\footnote{Based on observations obtained with: (1) the NASA/DLR Stratospheric Observatory for Infrared Astronomy (SOFIA). SOFIA is jointly operated by the Universities Space Research Association, Inc. (USRA), under NASA contract NAS2-97001, and the Deutsches SOFIA Institut (DSI) under DLR contract 50 OK 0901 to the University of Stuttgart; and (2) the MMT Observatory on Mt. Hopkins, AZ, a joint facility of the Smithsonian Institution and the University of Arizona.}}

\author[0000-0002-1913-2682]{Michael S. Gordon}
\affiliation{Minnesota Institute for Astrophysics, School of Physics and Astronomy\\116 Church St SE, University of Minnesota, Minneapolis, MN 55455, USA}

\author[0000-0003-1720-9807]{Roberta M. Humphreys}
\affiliation{Minnesota Institute for Astrophysics, School of Physics and Astronomy\\116 Church St SE, University of Minnesota, Minneapolis, MN 55455, USA}

\author[0000-0002-8716-6980]{Terry J. Jones}
\affiliation{Minnesota Institute for Astrophysics, School of Physics and Astronomy\\116 Church St SE, University of Minnesota, Minneapolis, MN 55455, USA}

\author[0000-0002-2398-145X]{Dinesh Shenoy}
\affiliation{Minnesota Institute for Astrophysics, School of Physics and Astronomy\\116 Church St SE, University of Minnesota, Minneapolis, MN 55455, USA}

\author{Robert D. Gehrz}
\affiliation{Minnesota Institute for Astrophysics, School of Physics and Astronomy\\116 Church St SE, University of Minnesota, Minneapolis, MN 55455, USA}

\author{L. Andrew Helton}
\affiliation{USRA-SOFIA Science Center, NASA Ames Research Center, Moffett Field, CA 94035, USA}

\author[0000-0001-9910-9230]{Massimo Marengo}
\affiliation{Department of Physics and Astronomy, Iowa State University, Ames, IA 50011, USA}

\author{Philip M. Hinz}
\affiliation{Department of Astronomy/Steward Observatory, University of Arizona, 933 N. Cherry Avenue, Tucson, AZ 85721, USA}

\author{William F. Hoffman}
\affiliation{Department of Astronomy/Steward Observatory, University of Arizona, 933 N. Cherry Avenue, Tucson, AZ 85721, USA}

\correspondingauthor{Michael S. Gordon}
\email{gordon@astro.umn.edu}

\begin{abstract}
  New MMT/MIRAC (9--11~\micron), \SOFIA/FORCAST (11--37~\micron), and Herschel/PACS (70 and 160~\micron) infrared (IR) imaging and photometry is presented for three famous OH/IR red supergiants (NML~Cyg, VX~Sgr, and S~Per) and two normal red supergiants (RS~Per and T~Per). We model the observed spectral energy distributions (SEDs) using radiative transfer code DUSTY. Azimuthal average profiles from the \SOFIA/FORCAST imaging, in addition to dust mass distribution profiles from DUSTY, constrain the mass-loss histories of these supergiants. For all of our observed supergiants, the DUSTY models suggest that constant mass-loss rates do not produce enough dust to explain the observed infrared emission in the stars' SEDs. Combining our results with \cite{shenoy2016} (Paper~I) we find mixed results with some red supergiants showing evidence for variable and high mass-loss events while others have constant mass loss over the past few thousand years.
\end{abstract}


\section{Introduction}
The evolution and fate  of massive stars depends on mass loss and their mass loss histories. The majority of massive stars ($\ge$~9~M$_{\odot}$) will pass through the red supergiant (RSG) stage, long recognized as an important end product of stellar evolution. Recently, \cite{smartt2009} and \cite{smartt2015} have suggested that RSGs with initial masses greater than 18~M$_{\odot}$ do not explode as supernovae, but may evolve back to warmer temperatures before the terminal explosion or collapse directly to black holes. The RSG stage is also a high mass losing stage, and to what extent mass loss can affect the terminal state of the RSGs is now an open question. Even though the mass-loss mechanism for RSGs is still debated, we can measure the mass lost from the thermal infrared (IR) emission from dust in the circumstellar ejecta surrounding the RSGs.

In our first paper (\citealt{shenoy2016}; hereafter, Paper~I), we examined the cold dust in the mid- to far-IR and the mass-loss histories of the famous hypergiants $\mu$~Cep, VY~CMa, IRC~+10420, and $\rho$~Cas, whose mass-loss rates are among the highest observed. In this paper, we present similar observations of three strong IR and maser sources, the OH/IR red supergiants NML~Cyg, VX~Sgr, and S~Per, plus the normal red supergiants RS~Per and T~Per. OH/IR stars, characterized by strong winds and OH maser emission, are bright IR sources due to thermal dust emission by their own circumstellar ejecta.  The more typical red supergiants, without OH or H$_2$O maser emission, also show high mass-loss rates that increase as a function of luminosity \citep{reimers1975,dejager1988,mauron2011}. In this study, we analyze the mass loss in these five red supergiants through observations in the mid-IR with \SOFIA/FORCAST \citep{herter2012} $11-37$~\micron\ imaging, combined with publicly-available \textit{Herschel}\footnote{\textit{Herschel} is an ESA space observatory with science instruments provided by the European-led Principal Investigator consortia and with important participation from NASA. The \textit{Herschel} data used in this paper are from the Level 2 (flux-calibrated) images provided by the Herschel Science Center via the NASA/IPAC Infrared Science Archive (IRSA), which is operated by the Jet Propulsion Laboratory, California Institute of Technology, under contract with NASA.} \citep{pilbratt2010} PACS \citep{poglitsch2010} images. We also include sub-arcsecond resolution $8-12$~\micron\ observations of NML~Cyg \citep{schuster2009}, S~Per, and T~Per made with MMT/MIRAC \citep{hoffmann1998,hinz2000}.

Finally, we present spectral energy distribution (SED) models from the radiative transfer code DUSTY \citep{ivezic1997}. These SED models, in combination with azimuthal profiles from FORCAST and MIRAC, provide estimates on mass-loss rates, ejecta dust temperatures, and mass-loss histories.

\section{Observations and Data Reduction}
\subsection{\SOFIA/FORCAST: Far-IR Imaging ($11-37$~\micron)}\label{sec:imaging}
The targets were observed with \SOFIA/FORCAST during Cycles 3 \& 4 (OBS IDs: 03\_0082, 04\_0013; PI: R.~M.~Humphreys).  FORCAST is a dual-channel mid-IR imager covering the 5 to 40~\micron\ range.  Each channel uses a 256$\times$256 pixel blocked-impurity-band (BiB) array and provides a distortion-corrected 3\farcm2$\times$3\farcm2 field of view with a scale of 0\farcs768~pix$^{-1}$.  FORCAST achieves near-diffraction limited imaging, with a PSF FWHM of $\sim$3\farcs7 in the longest filters.  We elected to image in single-beam mode to maximize throughput.  The observations were obtained  in the standard two-position chop-and-nod mode with the direction of the nod matching the direction of the chop (NMC).  The data were reduced by the SOFIA Science Center using the FORCAST Redux pipelines version 1.0.3 (S~Per), 1.0.5 (VX~Sgr), 1.0.7 (NML~Cyg), and 1.1.0 (RS~Per, T~Per).  After correction for bad pixels and droop effects, the pipeline removed sky and telescope background emission by first subtracting chopped image pairs and then subtracting nodded image pairs.  The resulting positive images were aligned and merged.  The details of the FORCAST pipeline are discussed in the Guest Investigator Handbook for FORCAST Data Products, Rev.\ A3.\footnote{\footnotesize{Available at \url{https://www.sofia.usra.edu/researchers/data-products}}}

Bright point sources cause cross-talk in the horizontal direction on the FORCAST array.  To mitigate this effect, chop angles were selected so that the cross-talk pattern from one chop position did not overlap with the other chop position.  Additionally, the FORCAST pipeline applies a correction that reduces the effect, although some of the pattern remains for some targets.  The effect is strongest for the brightest IR targets, NML~Cyg and VX~Sgr; it is less so for S~Per and is not present in the images of RS~Per or T~Per.   However, one effect that may appear in some of the fainter targets (especially T~Per) is possible coma introduced from the NMC chopping pattern.\footnote{\footnotesize{See description of optical aberrations in \S1.3.1 of the \SOFIA\ Observer's Handbook, available at \url{https://www.sofia.usra.edu/science/proposing-and-observing/sofia-observers-handbook-cycle-6}}}  This effect may explain the asymmetries in the surface brightness profile of T~Per, shown in Figure~\ref{fig:TPerRad} and discussed in \S\ref{sec:rsper}.

For each of the stars, observations of the asteroid 2~Pallas were used for PSF calibration in the same filters with the four-position slide in either the mirror position (for the short wavelength channel) or the open position (for the long wavelength channel).  The color temperature of the asteroid Pallas ($\sim160$~K) is far less than the effective temperatures of the target stars ($\gtrsim3200$~K).  This color difference is not ideal for a PSF calibrator, since the cooler source will peak at longer wavelengths, possibly resulting in a broader profile.  However, Pallas was the only source observed in each \SOFIA\ cycle under the same conditions and at each wavelength studied in this work.  We analyzed another calibrator ($\alpha$~Aur) at 11.1 and 31.5~\micron\ and measured a similar FWHM at each wavelength.  We present the profiles of Pallas in the figures below for consistency, acknowledging the possibility that we have overestimated the size of the PSF at the longer wavelengths.

The FORCAST pipeline coadds the merged images.  We use the standard deviation of the mean of fluxes extracted from the merged images (prior to coadding) as the 1-$\sigma$ uncertainty of the fluxes in the coadded images of each of our targets.   This uncertainty is negligible compared to the 6\% uncertainty that we adopt for the flux calibration, per the GI Handbook \S~4.1 \citep{herter2013}.  The band-passes of the selected FORCAST filters are such that only small color corrections are required.  Based on the $F_{\nu}\propto\nu^{2}$ spectral shapes of our targets in the relevant ranges, we have applied color corrections of 1.004, 1.071, 1.004, 1.044, 1.025, and 1.025 to fluxes extracted from the F111, F197, F253, F315, F348, and F371 images respectively.  Aperture photometry was performed using the open-source \textit{Astropy} \citep{astropy2013} affiliated \textit{photutils}\footnote{\footnotesize{\textit{photutils} provides tools for detecting and measuring the photometry of astronomical sources. The software is still in development, with documentation available at \url{https://photutils.readthedocs.io/}}} package. Apertures span between 15$\arcsec$ and 20$\arcsec$, chosen to encompass the extended emission around each object. FORCAST photometry is reported in Table~\ref{tab:phot} and included in the SEDs in \S\ref{sec:results}. Photometric error is reported as measured uncertainty in the sky background apertures.

\begin{deluxetable}{llllll}
  \tablecaption{Summary of Observations\label{tab:obs}}
  \tablecolumns{6}
  \tablenum{1}
  \tablehead{
    \colhead{Target} & \colhead{Instrument} & \colhead{Date} & \colhead{Filter\tablenotemark{a}} & \colhead{Int Time} & \colhead{PSF~FWHM\tablenotemark{b}} \\
    \colhead{} & \colhead{} & \colhead{(UT)} & \colhead{(\micron)} & \colhead{(s)} & \colhead{($\arcsec$)}}
  \startdata
  VX~Sgr & FORCAST & 2015 06 13 & F111, F197, F253   &  77, 81, 232 & 2.8, 2.9, 2.8 \\
  & & &  F315, F348, F371 & 224, 354, 496  & 3.1, 3.4, 3.6 \\
  S~Per & MIRAC4    & 2006 11 05 & 8.9, 9.8   & 420, 60  & 0.3, 0.4\\
  & FORCAST & 2015 02 04 & F197, F253   & 164, 374  & 2.7, 3.2 \\
  &  &  & F315, F348, F371 & 374, 705, 1430 & 3.4, 3.6, 3.8 \\
  RS~Per & FORCAST  & 2016 02 18  &  F197, F315, F371 & 29, 270, 1062  & 2.6, 3.1, 3.6 \\
  T~Per & MIRAC4 & 2009 10 02 & 8.9, 9.8 & 50, 70 & 0.32, 0.35 \\
  & FORCAST & 2016 09 17 & F197, F315 & 160, 484 & 2.6, 3.1 \\
  NML~Cyg\tablenotemark{c} & MIRAC3 & 2006 07 23 & 8.9, 9.8, 11.9 & 260, 260, 220 & 0.3, 0.4, 0.6 \\
  & FORCAST & 2015 09 11 & F197, F253  &  65, 233 & 2.6, 2.8 \\
  & & &  F315, F348, F371 & 320,  582, 342  & 3.2, 3.4, 3.6 \\
  \enddata
  \tablenotetext{a}{The effective wavelengths of the \SOFIA/FORCAST filters are:  F197 = 19.7~\micron, F253 = 25.3~\micron, F315 = 31.5~\micron, F348 = 34.8~\micron, F371 = 37.1~\micron.}
  \tablenotetext{b}{For \SOFIA, FWHM measured on PSF calibrator source 2~Pallas. Different cycles on \SOFIA\ may have very slight differences in apparent spatial resolution for the same filter sets. For MIRAC, FWHM measured on $\beta$~And.}
    \tablenotetext{c}{NML~Cyg MIRAC3/BLINC observations originally published in \cite{schuster2009}.}
\end{deluxetable}

\begin{deluxetable}{lrrrrrrrrrr}
  \tabletypesize{\scriptsize}
  \tablecaption{New Mid-Infrared Photometry\label{tab:phot}}
  \tablecolumns{11}
  \tablenum{2}
  \tablehead{
    \colhead{Name} & \colhead{8.8~\micron\tablenotemark{a}} & \colhead{9.8~\micron\tablenotemark{a}} & \colhead{11.1~\micron\tablenotemark{b}} & \colhead{19.7~\micron\tablenotemark{b}} & \colhead{25.3~\micron\tablenotemark{b}} & \colhead{31.5~\micron\tablenotemark{b}} & \colhead{34.8~\micron\tablenotemark{b}} & \colhead{37.1~\micron\tablenotemark{b}} & \colhead{70~\micron\tablenotemark{c}} & \colhead{160~\micron\tablenotemark{c}} \\
    \colhead{} & \colhead{(Jy)} & \colhead{(Jy)} & \colhead{(Jy)} & \colhead{(Jy)} & \colhead{(Jy)} & \colhead{(Jy)} & \colhead{(Jy)} & \colhead{(Jy)} & \colhead{(Jy)} & \colhead{(Jy)}
  }
  \startdata
  \objectname{VX Sgr} & \nodata & \nodata & $3740\pm98$ & $2250\pm91$ & $1400\pm85$ & $1090\pm69$ & $835\pm75$ & $697\pm10$ & $153\pm32$ & $23\pm11$ \\
  \objectname{S Per} & $316\pm52$ & $340\pm48$ & \nodata & $342\pm28$ & $187\pm24$ & $146\pm15$ & $109\pm9.4$ & $97\pm9.6$ & $21\pm4.1$ & $3\pm1.2$ \\
  \objectname{RS Per} & \nodata & \nodata & \nodata & $82\pm4.8$ & \nodata & $21\pm2.0$ & \nodata & $15\pm4.1$ & \nodata & \nodata \\
  \objectname{T Per} & $8.8\pm1.4$ & $11.4\pm3.1$ & \nodata & $9.7\pm1.2$ & \nodata & $7.2\pm2.9$ & \nodata & \nodata & \nodata & \nodata \\
  \objectname{NML Cyg}\tablenotemark{d} & $3735\pm63$ & $3780\pm160$ & \nodata & $4868\pm41$ & $3930\pm328$ & $3626\pm110$ & $2849\pm123$ & $2809\pm287$ & $652\pm96$ & $116\pm30$ \\
  \enddata
  \tablenotetext{a}{MMT/MIRAC}
  \tablenotetext{b}{SOFIA/FORCAST}
  \tablenotetext{c}{HERSCHEL/PACS}
  \tablenotetext{d}{NML~Cyg MIRAC photometry originally presented in \cite{schuster2009}.}
\end{deluxetable}

\subsection{Adaptive Optics Mid-IR Imaging ($8-10$~\micron)}
NML~Cyg, S~Per, and T~Per were observed with the mid-IR adaptive optics system on the MMT using the Mid-Infrared Array Camera and Bracewell Infrared Nulling Cryostat \citep[MIRAC3/4/MIRAC-BLINC;][]{hoffmann1998, hinz2000, skemer2008}.  The observations of NML~Cyg are described in \cite{schuster2009}, and discussed here in \S\ref{sec:nmlcyg}.  S~Per was observed on UT~2006~Nov~05 and T~Per on UT~2009~Oct~02 at 8.9 and 9.8~\micron.  MIRAC achieved Strehl-ratios close to 0.95, providing diffraction-limited imaging and stable PSFs \citep[e.g.][]{biller2005}.   MIRAC4 employed a Si:As array with 256$\times$256 pixels, and observations were made with a standard chop-and nod sequence to remove IR background emission.  Cross-talk in the array electronics introduced faint artifacts in the horizontal and vertical directions which is not completely removed by chop-and-nod subtraction. As described in Paper~I, the horizontal cross-talk is mitigated during the data reduction with a code from \cite{skemer2008}.  For consistency with the FORCAST photometry, we perform aperture photometry on the MIRAC images using \textit{photutils}. We report the results in Table~\ref{tab:phot}, include the photometry in the SEDs below, and as input to DUSTY.

\subsection{IRAS, AKARI, WISE, and ISO-SWS ($2-100$~\micron)}
To populate the mid-IR SEDs, we include IRAS photometry (and AKARI photometry when available) from point-source catalogs in the literature for RS~Per and VX~Sgr \citep{smith2004}, S~Per and T~Per \citep{abrahamyan2015}, and NML~Cyg \citep{schuster2007}. The \cite{abrahamyan2015} catalog cross-correlates IRAS point sources with WISE, the latter of which presents some issues due to its large beam-size \citep[up to 12\arcsec\ at 22~\micron;][]{wright2010}. For stars embedded in nebulosity or crowded fields, the WISE photometry can be systematically too bright.

Additionally, optical photometry is compiled from the Extended Hipparchos Compilation catalog \citep[XHIP;][]{anderson2012}, the SKY2000 Master Catalog \citep{myers2015}, or the AAVSO Photometric All Sky Survey \citep[APASS;][]{henden2016} (see SEDs in \S\ref{sec:results}).  These optical data, as well as the published photometry from IRAS and AKARI, are dereddened using the extinction law from \cite{odonnell1994}. The values for interstellar extinction $A_V$ chosen for each source are listed in Table~\ref{tab:models} and in the SED captions below.

We also compile spectra from ISO-SWS \citep{degraauw1996} for all targets except T~Per.  S~Per and RS~Per spectra are from the Japanese guaranteed observing time program REDSTAR1 \citep[PI T. Tsuji;][]{aoki1998}, and NML~Cyg and VX~Sgr were observed with the AGBSTARS program \citep{justtanont1996,speck2000}.  The color and extinction-corrected spectra are displayed in the SEDs below and are provided as near- to mid-IR photometric input to DUSTY.

\subsection{Herschel/PACS (70, 160~\micron)}
We also include in our analysis the publicly-available 70 and 160~\micron\ observations made with Herschel/PACS. VX~Sgr, NML~Cyg, and S~Per were observed as part of the Herschel key program Mass-loss of Evolved StarS \citep[MESS;][]{groenewegen2011}. The Herschel Interactive Processing Environment \citep[HIPE;][]{ott2010}\footnote{\footnotesize{HIPE is a joint development by the Herschel Science Ground Segment Consortium, consisting of ESA; the NASA Herschel Science Center; and the HIFI, PACS, and SPIRE consortia.}} was used to download the images, but photometry was performed using \textit{photutils} for consistency with the \SOFIA\ images. Apertures span between 45 and 70$\arcsec$ to encompass extended emission around each object.  Since the PACS pixels are large on sky, we did not have enough pixels in traditional sky annuli to model the background. Instead, we first mask the star and its nebulosity, and then model the background across the field of view as a 2D polynomial. For each of the PACS fields, these background models were fairly flat but had high RMS variation. As summarized in Table~\ref{tab:phot}, this uncertainty was as high as $\sim40\%$ for VX~Sgr and S~Per.

The width of the PACS bandpasses requires color corrections to be applied to the 70 and 160~\micron\ photometry from the images. In Paper~I, the necessary corrections were estimated by convolving the ``blue'' (70~\micron) filter response functions to the sources' ISO LWS spectra. However, lacking spectra for all of the sources in this work, we instead fit the mid- to far-IR photometry from \SOFIA\ and IRAS with a power-law of the form $F_\nu = \nu^\beta$ to represent the targets' SEDs at the PACS wavelengths. The results are modest corrections of 1.003 and 1.04 for the two bandpasses. PACS photometry is reported in Table~\ref{tab:phot} and included in the SEDs in \S\ref{sec:results}. Photometric error is reported as measured uncertainty in the sky background models.

\section{Results \& Discussion}\label{sec:results}
\subsection{DUSTY modeling}\label{sec:dusty}
To estimate the mass-loss rates, mass-loss histories, and dust density distributions, we used the DUSTY radiative-transfer code \citep{ivezic1997} to model the observed SEDs and azimuthal average intensity profiles at each of the MIRAC and FORCAST wavelengths, in a manner similar to that used in Paper~I.  DUSTY solves the 1D radiative-transfer equation for a spherically-symmetric dust distribution around a central source. We provide as input the chosen optical properties, chemistry, size distribution of the dust grains, and a dust temperature, which fixes the inner boundary of the surrounding dust shell (the dust condensation radius, $r_1$). We generate a grid of models for each star with fixed stellar effective temperatures based on the spectral type of each target, fixed shell extent ($1000\times r_1$), and fixed dust condensation temperature (1000 K).  Our grid consists of varying optical depths of the circumstellar material ($0.01 < \tau_V < 50$) and different dust density distribution functions, described below. For a given set of inputs, DUSTY outputs a model SED and radial profiles of the dust shell at requested wavelengths. 

As noted in Paper~I, the spherical symmetry assumed by DUSTY fails to model the azimuthal complexities observed in the asymmetric outflows of massive stars such as VY~CMa \citep{smith2001,humphreys2005,humphreys2007,shenoy2013} and IRC~+10420 \citep{humphreys1997,tiffany2010,shenoy2015}. However, DUSTY allows for a consistent analysis of the dust, SEDs, and intensity profiles of the targets in this work and those in Paper~I.

At a given wavelength, an output optical depth $\tau_\lambda$ from the model, and thus its grain opacity $\kappa_\lambda$, specifies the dust mass density $\rho\left(r\right)$ throughout the shell. If we assume a constant expansion rate $v_{exp}$ of the outflowing dust shell, following the arguments set forth in Paper~I, we can estimate the mass loss rate as:
\[
\dot{M}\left(t\right)=g_d \,4\pi r^2\,\rho\left(r\right)\,v_{exp}
\]
where radius $r$ is a probe on timescale $t$ since $r=v_{exp}\, t$, and $g_d$ is the gas-to-dust ratio. For consistency with Paper~I, we assume $g_d =$~100:1 \citep{knapp1993}; however, this can be as high as 200:1 for supergiants \citep{decin2006,mauron2011}. For the dust optical properties, we use the ``cool'' circumstellar silicates from \citet{ossenkopf1992}, and assume the grain radii follow a Mathis, Rumpl, Nordsieck (MRN) size distribution $n\left(a\right) \propto a^{-3.5}\,da$ \citep{mathis1977} with $a_{min} = 0.005$~\micron\ and $a_{max} = 0.25$~\micron.

In general, the mass density distribution of the outflows can be modeled with DUSTY as a power-law $\rho\left(r\right)\propto r^{-q}$.  An index of $q=2$ is the case of constant mass-loss rate and constant expansion velocity for the shell, while $q<2$ indicates a gradual decline in the mass-loss rate over the dynamical age of the expanding shell.  A steeper power-law index $q>2$ represents a mass distribution with more recent high mass loss, and less significant mass loss in the past.  For each of our targets, the fundamental research question is how well the stars' SEDs and radial profiles in the mid-IR can be modeled with a constant mass-loss rate scenario in DUSTY.

For each of the targets in our sample, we perform three Monte Carlo experiments.  In the first, we force DUSTY to use the constant mass-loss rate distribution $\rho\left(r\right)\propto r^{-2}$, and by varying the optical depth of the CS material, recover the best-fitting, constant mass-loss SED in the near- to mid-infrared.  For the second set of simulations, we allow the power-law index of the mass distribution to vary between 1 and 3 with a step size of 0.2, deliberately excluding the $r^{-2}$ case, while also allowing the optical depth to vary.  For both set of DUSTY models, we evaluate the best fit based on a reduced $\chi^2$ measurement of the extinction-corrected SED and the DUSTY output spectrum.  We then compare the DUSTY-predicted intensity profiles to the observed radial profiles in the \SOFIA\ wavelengths (and MIRAC, when available).  The image and profile models output from DUSTY do not account for the optics of the telescopes, so the intensity profiles are convolved with an azimuthal average of the PSF and are displayed in the figures below.

For the third and final set of DUSTY models, we select the best-fitting $r^{-2}$ model, and re-run DUSTY with those same parameters, this time enhancing the dusty density profile by a factor of ten at 50 condensation radii ($50\times r_1$). An example model is shown in Figure~\ref{fig:enhanced}.  These ``enhanced,'' piecewise-defined models explore the possibility of an extreme mass-loss event in a star's past, similar in some respects to the models from the second experiment described above where the density distribution can be shallower than $r^2$. The latter DUSTY models imply a smoothly-changing mass-loss rate over the lifetime of the star, whereas the ``enhanced'' models simulate a single eruptive event in the mass-loss history. Similar piece-wise defined density profiles were used in Paper~I to model the SED of IRC~+10420. Though a factor of ten enhancement in mass-loss rate is likely an extreme case, we apply this model to explore how well the scenario of a constant mass-loss rate with a single one-off eruptive event reproduces the observed IR SED of our target stars.

Finally, we can estimate an average mass-loss rate for the non-constant mass-loss models ($q\neq2$) by integrating the density distribution $\rho\left(r\right)$ and multiplying by the gas-to-dust mass ratio (100:1) to compute the total mass of the shell $M$.  We assume an average expansion velocity to estimate the dynamical age of the shell $\Delta t=r_2/v_{exp}$ where $r_2$ is the outer radius of the shell predicted by a given model. The expansion velocity, $v_{exp}$, is assumed to be 25~km~s$^{-1}$ unless specified in the sections for individual stars below. The average mass-loss rate is then $\langle\dot{M}\rangle=M/\Delta t$. The specific parameters for the DUSTY models for each target in our program, as well as the output DUSTY models and computed average mass-loss rates, are summarized in Table~\ref{tab:models}, and the best-fitting SEDs to the observed photometry are shown in the figures below.

Note that in Table~\ref{tab:models}, the first row for each target star represents the best-fitting DUSTY simulations forced to evaluate the models in the constant mass-losing, $r^{-2}$ dust profile case.  The second row represents the best-fitting SED with non-constant mass-loss.  The columns on the left reflect the input values, and the right-hand columns are the recovered output parameters from the best-fitting models for each target and each simulation set (constant vs.\ non-constant mass-loss rates). We do not include the enhanced, piecewise-defined models here, since the parameters were fixed to the $r^{-2}$ model for each star.  Throughout the text we will refer to the three different models as constant ($r^{-2}$), non-constant ($r^{-q},\,q\neq2$), and enhanced ($r^{-2}$, e) mass-loss rates.

\begin{deluxetable}{llll|rcl}
  \tablecaption{DUSTY Model Parameters and Mass-Loss Rates\label{tab:models}}
  \tablecolumns{7}
  \tablenum{3}
  \tablehead{
    \colhead{Model} & \colhead{A$_{\mathrm{V}}$\tablenotemark{a}} & \colhead{T$_{\mathrm{eff}}$} & \colhead{T$_{\mathrm{dust}}$\tablenotemark{b}} & \colhead{$\tau_V$} & \colhead{r$_1$} & \colhead{$\dot{\mathrm{M}}$\tablenotemark{c}} \\
    \colhead{} & \colhead{ } & \colhead{K} & \colhead{K} & \colhead{} & \colhead{AU} & \colhead{$\dot{\mathrm{M}}_\odot$/yr}}
  \startdata
  \multicolumn{4}{c}{Inputs} & \multicolumn{3}{c}{Outputs}  \\
  \hline
  \multicolumn{2}{c}{\bf VX~Sgr} \\
  $r^{-2.0}$ & \multirow{2}{*}{2.0} & \multirow{2}{*}{3200} & \multirow{2}{*}{1000} & 6.7 & 86 & $4.5 \times 10^{-5}$ \\
  $r^{-1.6}$ & & & & 3.7 & 76 & $2.2 \times 10^{-5}$ \\
  \hline
  \multicolumn{2}{c}{\bf S~Per} \\
  $r^{-2.0}$ & \multirow{2}{*}{3.1} & \multirow{2}{*}{3500} & \multirow{2}{*}{1000} & 1.4 & 45 & $2.6 \times 10^{-5}$ \\
  $r^{-1.6}$ & & & & 1.2 & 43 & $2.4 \times 10^{-5}$ \\
  \hline
  \multicolumn{2}{c}{\bf RS~Per} \\
  $r^{-2.0}$ & \multirow{2}{*}{1.7} & \multirow{2}{*}{3600} & \multirow{2}{*}{1000} & 0.3 & 50 & $3.5 \times 10^{-5}$ \\
  $r^{-1.6}$ & & & & 0.3 & 50 & $3.9 \times 10^{-5}$ \\
  \hline
  \multicolumn{2}{c}{\bf T~Per} \\
  $r^{-2.0}$ & \multirow{2}{*}{2.1} & \multirow{2}{*}{3700} & \multirow{2}{*}{1000} &  0.1 & 30 & $7.5 \times 10^{-6}$ \\
  $r^{-1.6}$ & & & & 0.1 & 31 & $8.1 \times 10^{-6}$ \\
  \hline
  \multicolumn{2}{c}{\bf NML~Cyg} \\
  $r^{-2.0}$ & \multirow{2}{*}{4.0} & \multirow{2}{*}{3300} & \multirow{2}{*}{1000} & 41 & 133 & $4.8 \times 10^{-4}$ \\
  $r^{-1.8}$ & & & & 37 & 128 & $4.2 \times 10^{-4}$ \\
  \enddata
  \tablenotetext{a}{DUSTY output models are fit to extinction-corrected SEDs with these values of A$_V$.}
  \tablenotetext{b}{Dust temperature at condensation radius, r$_1$.}
  \tablenotetext{c}{$\dot{\mathrm{M}}$ is computed as an average mass-loss rate over the lifetime of the shell. The outflow velocity is assumed to be 25~km/s unless noted in the sections for the individual stars.}
\end{deluxetable}

\subsection{VX~Sgr}
VX~Sgr has a marginally-resolved, nearly symmetric extended circumstellar envelope in its HST visual images \citep{schuster2006}. Additionally, \cite{vlemmings2005} has identified a dipole magnetic field in its ejecta mapped by its H$_{2}$O masers, which may be a clue to its mass loss mechanism.  VX~Sgr is also a semi-regular variable that behaves like a fundamental mode pulsator (i.e.\ a Mira variable), which is rare for such a luminous star. It has been observed to vary by several magnitudes with corresponding changes in its apparent spectral type from M4 to M10.

During one of its Mira-like episodes, \cite{humphreys1972} noted a decline of $\sim0.5$~mag out to 10~\micron\ over a few months. Therefore, we have chosen optical photometry to align in light-curve phase with the 2MASS and IRAC photometry compiled in \cite{smith2004}.  To constrain the 2--10~\micron\ regime of the SED, we also include the photometric average over the 3.6-yr cycle of the COBE DIRBE project \citep{price2010}.

The SED is shown in Figure~\ref{fig:VXSgrSED}, with observed data plotted as open symbols and extinction-corrected photometry in solid. The constant mass-loss DUSTY model is overplotted with a dashed blue line, and the best-fitting power-law model with index $q=1.6$ is shown in dashed-dotted green. Note that both models fit the 10~\micron\ silicate feature in the ISO spectrum, but the $q=1.6$ model better simulates the cool thermal dust emission out to 100~\micron.  However, the observed IR flux from $\sim30-70$~\micron\ falls in between the two models. Displayed in dotted red is the ``enhanced'' DUSTY model, which is the $r^{-2.0}$ constant mass-loss model with a factor of 10 enhancement in dust density at $50\times r_1$.  This model appears to over-estimate the thermal dust emission, implying too much dust is produced to match the observations of VX~Sgr.

The derived mass-loss rates for the models are summarized in Table~\ref{tab:models}.  The outflow velocity $v_{exp}$ adopted for this calculation is 24.3~km/s from the AGB/supergiant CO-line survey by \cite{debeck2010}.  The mass-loss rates from DUSTY ($2-5\times10^{-5}\,\mathrm{M}_\odot$/yr) are somewhat lower than the measurements from CO-line profiles \citep[$6.1\times10^{-5}\,\mathrm{M}_\odot$/yr;][]{debeck2010}.  However, the most obvious explanation for this is due to the assumed gas-to-dust ratio. Where we have assumed 100:1 for consistency with Paper~I, \cite{debeck2010} allow the gas-to-dust ratio to vary when fitting the observed outflow velocities \citep[using GASTRoNOoM;][]{decin2006}.  Mass-loss rates scale linearly with the gas-to-dust ratio, so if we had applied a ratio of 200:1, perhaps more appropriate for RSGs \citep{decin2006,mauron2011}, our estimated mass-loss rate would be more consistent with the derived measurement from CO-line profiles.

In Figure~\ref{fig:VXSgrRad}, we compare the observed radial profiles to the PSF calibrator (2 Pallas) and the DUSTY output image models.  The ejecta around VX~Sgr is only marginally resolved above the PSF at 19.7~\micron, but the envelope is more easily distinguished from the PSF at longer wavelengths.  The DUSTY model profiles are convolved with the PSF in each band, and we note that the constant mass-loss rate model aligns more closely with the observed surface brightness profiles of VX~Sgr.  Both the shallower $r^{-1.6}$ and enhanced models over-estimate the amount of thermal dust emission observed in the FORCAST images.

The azimuthal profiles combined with the SED modeling suggest that the mass-loss rate of VX~Sgr is fairly constant with perhaps a period of elevated mass-loss in the past, as illustrated by the infrared excess emission in the $20-160$~\micron\ photometry. The model intensity profiles, though, all predict a higher surface brightness for the extended emission than what was actually observed at SOFIA wavelengths (Figure~\ref{fig:VXSgrRad}). One possible explanation for not observing this emission is that VX~Sgr has experienced a sudden decline in mass-loss rate in very recent times. Exploring this possibility is beyond the scope of this paper, but we plan to make high-resolution $5-12$~\micron\ observations using LMIRCam and NOMIC on the LBT \citep{skrutskie2010,hoffmann2014}. At FWHM spatial resolutions of 0\farcs12 and 0\farcs29 at 5 and 12~\micron, respectively, we can explore the dust shell at $\sim200$~AU scales and combine these observations with our SOFIA data and DUSTY modeling.

We also note that although the envelope is spherically-symmetric in HST optical images \citep{schuster2006}, the H$_{2}$O masers around VX~Sgr appear to align with the equatorial plane of the star's dipole magnetic field. \cite{vlemmings2005} suggest that this alignment could create an overdensity in the circumstellar material in this plane, as modeled by \cite{matt2000}.  While we do not see evidence for asymmetry in the FORCAST images, we note again that DUSTY assumes spherical symmetry in its models.  As discussed further with S~Per and NML~Cyg below, it is likely that DUSTY may fail to accurately model stars with known asymmetric outflows and profiles.




\subsection{S~Per}\label{sec:sper}
S Per (Sp.\ Type M3-4e~Ia) is an OH/IR source and a member of the Per~OB1 association \citep{humphreys1978}, with a distance of 2.3$\pm$0.1~kpc as determined by VLBI H$_{2}$O maser astrometry \citep{asaki2010}.  \citet{schuster2006} present \emph{HST} images showing that the star is embedded in an elongated circumstellar envelope with a position angle of $\sim20\degr$ E of N with a FWHM of $\sim0\farcs1$ (240~AU).  \cite{schuster2006} speculated that the shape could be due to bipolarity in the star's ejecta or a flattened circumstellar halo, and they note this elongated structure is also seen in OH and H$_{2}$O maser observations on the same scale and with similar orientation \citep{richards1999,vlemmings2001}. Fitting elliptical Gaussians to S~Per's MIRAC4 images yields a mean position angle of 19$\degr$ $\pm$ 2$\degr$ E of N, matching closely the orientation seen in the HST images.

The observed SED is shown in Figure~\ref{fig:SPerSED} along with the three DUSTY models. Both the shallower $q=1.6$ and enhanced DUSTY models accurately reconstruct the near- to mid-IR flux, while the constant mass-loss rate model underestimates the thermal dust emission. We note here one possible complication in our analysis. DUSTY simulations assume spherical symmetry in CS material, which could lead to underestimating the density, and thus optical depth, of the ejecta relative to the observed compact envelope seen in the \cite{schuster2006} WFPC2 images of S~Per.  Our models for this star, then, may not best represent the stellar outflows and dusty envelope.

The observed azimuthal average radial profiles from \SOFIA/FORCAST for S~Per are presented in Figure~\ref{fig:SPerRad}. S~Per has resolvable extended emission above the PSF; however, the $q=2$ and $q=1.6$ profiles, once convolved with the large PSF beam of FORCAST, are virtually indistinguishable.  The enhanced DUSTY model, though, predicts too much emission close to the central star.

In Figure~\ref{fig:SPerRadMIRAC}, we illustrate the surface brightness profiles at higher spatial resolution with MIRAC. Here, the observed surface brightness profile is clearly resolved above the PSF at the shorter wavelengths; however, the DUSTY models underestimate the shape of the stellar envelope.  Adding a period of enhanced mass loss, the DUSTY model in dotted red, produces too much emission at the shorter wavelengths.  Unfortunately, then, the radial profile models do not provide any conclusive evidence that the mass-loss history of S~Per is constant vs.\ non-constant.  Note that the deviations from a smooth profile in the 31.5~\micron\ and the two MIRAC figures are due to the asymmetry in the outflows. From the SED, though, we glean that S~Per may have had a higher mass-loss rate in the past, but we acknowledge that DUSTY is not ideal for simulating stellar ejecta of stars with known bipolar/asymmetric envelopes.

As reported in Table~\ref{tab:models}, the two DUSTY models predict mass-loss rates between $\sim2-3\times10^{-5}\,\mathrm{M}_\odot$/yr. \cite{richards1999} summarizes results from previous literature to show a range of published mass-loss rates from as low as $7\times10^{-6}\,\mathrm{M}_\odot$/yr \citep[OH 1612~MHz;][]{jura1990} to as high as $2\times10^{-4}\,\mathrm{M}_\odot$/yr \citep[CO-line profiles;][]{knapp1985}. With such a large range of published values, each measuring mass-loss rates with a different observational technique, we can only conclude that we have derived a rate within published bounds.

\cite{fok2012} also performed DUSTY modeling on a number of Galactic RSGs, including S~Per, T~Per, and RS~Per.  However, they used a different mode, the ``dusty AGB'' radiatively-driven wind mode, and a higher gas-to-dust ratio of 200:1.  The radiatively-driven wind mode in DUSTY is provided for modeling AGB star envelopes and is not necessarily appropriate for RSGs \citep[e.g.,][]{heras2005}. \cite{groenewegen2012} analyzed the systematic difference in mass-loss rates computed using DUSTY in this mode as compared to the default where the user supplies the density distribution as a power-law function. \cite{groenewegen2012} found that the mass-loss rates computed with the radiatively-driven wind mode differ significantly from those obtained from the default mode in which the equation of radiative transfer alone is solved when applied in the context of RSGs.  Thus, our results, with fixed single-component power-law distributions, are not directly comparable to the \cite{fok2012} models.  Nonetheless, \cite{fok2012} yields a best-fitting model with $\dot{\mathrm{M}}=1\times10^{-5}\,\mathrm{M}_\odot$/yr. We note that \cite{gehrz1971} derived a mass-loss rate for S~Per of $2.7\times10^{-5}\,\mathrm{M}_\odot$/yr using an independent analysis of the 3.6--11.4~\micron\ SED. These results are consistent with our measurements.


\subsection{RS~Per and T~Per}\label{sec:rsper}
Departures from circular symmetry have been reported for both RS~Per and T~Per based on H-band interferometric imaging with CHARA by \cite{baron2014} at an angular scale of 1.3~mas, which the authors attribute to surface asymmetries or spots.  We do not see much evidence for asymmetry in the FORCAST images, and the azimuthal profiles in Figure~\ref{fig:RSPerRad} do not show significant excess emission above the PSF, though the angular scales for FORCAST are much larger than \cite{baron2014} observed with the CHARA array.  RS~Per has the 10~\micron\ silicate emission feature in its SED but is not a known maser source. It is likely a normal red supergiant that may just be entering a more active phase with enhanced mass loss, perhaps driven by surface activity like that seen in the OH/IR supergiants and VY~CMa. T~Per, another member of the Perseus~OB1 association, similarly shows no evidence for SiO maser emission \citep{jiang1999}. Both stars exhibit long-period variability of $\sim$4200 and 2500 days for RS~Per and T~Per, respectively \citep[AAVSO;][]{kiss2006}.

The SED of RS~Per is shown in Figure~\ref{fig:RSPerSED}. Both models fit the 10~\micron\ silicate feature, as well as the ISO spectrum, from $2-11$~\micron. However, the constant mass-loss, $q=2$ profile underestimates the flux for the mid- to far-IR at wavelengths larger than 20~\micron, while the shallower dust density distribution $q=1.6$ profile and the enhanced $r^{-2.0}$ profile better match the longer wavelength SED.  For T~Per, shown in Figure~\ref{fig:TPerSED}, both the constant mass-loss and enhanced models produce insufficient thermal dust emission at the longer wavelengths.

As summarized in Table~\ref{tab:models}, the mass-loss rates derived for RS~Per and T~Per are $4\times10^{-5}$ and $8\times10^{-6}\,\mathrm{M}_\odot$/yr, respectively.  \cite{fok2012} estimates a mass-loss rate of $3.0\times10^{-6}\,\mathrm{M}_\odot$/yr for RS~Per and $5\times10^{-7}\,\mathrm{M}_\odot$/yr for T~Per using DUSTY ``AGB mode.''  As discussed in \S\ref{sec:sper}, this radiatively-driven wind mode is less appropriate for RSGs, and so our results are not directly comparable.  Additionally, we note that the SED fits provided in their work do not extend longward of 30~\micron, so we cannot qualitatively gauge which SED modeling mode (our power-law profiles vs.\ their radiatively-driven wind models) would fit best with our new FORCAST photometry through 40~\micron.  Particularly in the case of T~Per, we note that this long-wavelength IR photometry is crucial in constraining the models.

While the $r^{-1.6}$ DUSTY model is clearly the better fit to the observed mid-IR SED, the radial profiles of RS~Per in Figure~\ref{fig:RSPerRad} reveal that the \SOFIA\ images lack the spatial resolution necessary to favor one model over the other as a better fit to the extended envelope emission. The radial profiles of T~Per in Figures~\ref{fig:TPerRad} and \ref{fig:TPerRadMIRAC} reveal a much clearer extended profile above the PSF at 31.5~\micron\ and around the 10~\micron\ silicate feature with MIRAC.  However, the two power-law models predict very similar output profiles.  The enhanced model accurately recovers the extended emission around the 10~\micron\ silicate feature, but the MIRAC images in the two other wavebands do not resolve emission extended above the PSF.

We note a curious ripple in the 31.5~\micron\ FORCAST profile in Figure~\ref{fig:TPerRad}.  As mentioned in \S\ref{sec:imaging}, some optical aberrations may be introduced to the images due to chopping patterns of the secondary mirror on \SOFIA.  One possible source of the profile shape could be coma effects that stretch out T~Per along one axis.  When generating azimuthally-averaged surface brightness profiles, this asymmetry would cause the profile to deviate from a smooth power-law.




\subsection{NML~Cyg}\label{sec:nmlcyg}
HST visual images of NML~Cyg revealed a peculiar bean-shaped asymmetric nebula only $\approx0\farcs2$ across and coincident with the distribution of its H$_{2}$O masers. \citet{schuster2006} showed that its circumstellar envelope is shaped by photodissociation from the powerful nearby association Cyg~OB2 inside the Cygnus~X superbubble which is relatively void of gas and dust.  This configuration allows the UV radiation from the numerous luminous hot stars in Cyg~OB2 to travel the $\approx80$~pc to NML~Cyg unimpeded.   Subsequent adaptive optics mid-IR imaging at 8.8, 9.8 and 11.7~\micron\ with MIRAC3 on the MMT \citep{schuster2009} spatially resolve the physical structures near the star ($\sim240$~AU) responsible for its 10~\micron\ silicate-absorption feature and an asymmetric excess at $0\farcs3 - 0\farcs5$ from the star due to thermal emission from hot dust. This excess is also oriented toward the Cyg~OB2 association and is attributed to the destruction of NML~Cyg's dusty wind by the hot stars in Cyg~OB2.

As illustrated in the SED in Figure~\ref{fig:NMLCygSED}, the 10~\micron\ silicate feature is seen in absorption, rather than emission.  Discussed in detail in \cite{schuster2006,schuster2009}, this absorption is due to NML~Cyg's thick circumstellar envelope obscuring the central star.  Indeed, the DUSTY models predict a large optical depth ($\tau_V>40$ for both models) to fit the silicate feature in absorption as well as the mid- to far-IR photometry.

At the FORCAST wavelengths, 19.7 to 37.1~\micron, NML~Cyg appears as a point source, with no evidence of cold dust hidden or protected from the UV radiation in Cyg~OB2.  Additionally, no preferential extension towards Cyg~OB2 is evident at the angular resolution of FORCAST.  The radial profiles in Figure~\ref{fig:NMLCygRad} do not show any obvious excess emission above the PSF. While we note that the shape of the observed azimuthal profile seems similar to the $r^{-2}$ model, the models seem to greatly over-predict the surface brightness flux for the large estimated optical depth.

As noted in \cite{schuster2009}, the MIRAC images do indeed appear asymmetric along the NW--SE axis.  For these images, we generate isophotes and separate our radial profiles into two axes.  The major axis (NW--SE) is shown with solid points in Figure~\ref{fig:NMLCygRadMIRAC}, and the minor axis (NE--SW) is shown with open circles.  Here, we see that the constant mass-loss, $q=2$, model does fit the observed major-axis brightness profile for the 9.8~\micron\ image, though the $q=1.8$ profile is similar in shape for all three MIRAC bands.  We note that DUSTY cannot take into account the external radiation field from Cyg~OB2, so the model profile shapes are not necessarily conclusive as to the mass-loss history of NML~Cyg.

NML~Cyg has one of the highest mass-loss rates of any red supergiant/hypergiant---from $6.4\times10^{-5}\,\mathrm{M}_\odot$/yr \citep{morris1983} to as high as $1.6\times10^{-4}\,\mathrm{M}_\odot$/yr \citep{lucas1992}.  We calculate even higher mass-loss rates from the DUSTY models, though, at $4-5\times10^{-4}\,\mathrm{M}_\odot$/yr with average outflow velocity 23~km~s$^{-1}$ \citep{schuster2009}.  It is likely, however, that the complex morphology of NML~Cyg cannot be well-modeled with a single-component power-law dust mass distribution, similar to the difficulties in modeling S~Per with its known asymmetric profile.


\section{Conclusions}\label{sec:conclusions}
With mid-infrared imaging from MMT/MIRAC and \SOFIA/FORCAST, we observed three OH/IR red supergiants, NML~Cyg, VX~Sgr, and S~Per and the normal red supergiants RS~Per and T~Per.  We present new photometry at 9--11~\micron\ with MIRAC, at 20--40~\micron\ with FORCAST, and at 70 and 160~\micron\ from the Herschel/PACS archive. These data, in combination with published optical and near- to mid-IR photometry, are used to constrain DUSTY model SEDs.

\textit{VX~Sgr}: Though a symmetric extended circumstellar envelope is resolved in HST images \citep{schuster2006}, we have only marginally resolved a cooler dust component at 19.7--37.1~\micron\ with FORCAST. From DUSTY, we conclude that the mass loss around VX~Sgr cannot necessarily be well modeled by smooth, constant mass loss.  The IR excess emission combined with DUSTY modeling show evidence for a higher mass-loss rate in the past for VX~Sgr with an average estimated mass-loss rate of $2\times10^{-5}\,\mathrm{M}_\odot$/yr for the $\rho\left(r\right)\propto r^{-1.6}$ model, and $5\times10^{-5}\,\mathrm{M}_\odot$/yr for the constant mass-loss case.

\textit{S~Per}: Azimuthal profiles of S~Per reveal an IR excess above the PSF emission from the central star in MIRAC imaging, and to a lesser extent in the FORCAST wavelengths. However, the radial profiles produced from DUSTY are not significantly different when convolved with the optics of \SOFIA.  Both the $r^{-1.6}$ dust density distribution model and the enhanced mass-loss model fit the near- to mid-IR SED out to 160~\micron, implying the possibility of a higher mass-loss rate in the past.  S~Per is known to posses an asymmetric outflow close to the central star and lack an extended spherical nebula; therefore, the 1D spherically-symmetric radiative-transfer code DUSTY may not be the most accurate method for reconstructing the mass-loss history of S~Per. We estimate an average mass-loss rate of $\sim2-3\times10^{-5}\,\mathrm{M}_\odot$/yr.

\textit{RS~Per} and \textit{T~Per}: SED models of both stars suggest that a constant mass-loss rate is insufficient to generate enough 20--40~\micron\ emission to match the observed mid-IR photometry.  The FORCAST images for both RSGs, and the MIRAC 10~\micron\ image for T~Per, show modest excess emission above the flux from the PSF.  The SED for RS~Per appears to be best fit with a shallow power-law distribution in dust density of $\rho\left(r\right)\propto r^{-1.6}$, suggesting it had a higher mass-loss rate in the past. Over the lifetimes of the observed dust shells, we estimate average mass-loss rates of $4\times10^{-5}\,\mathrm{M}_\odot/$yr for RS~Per and $8\times10^{-6}\,\mathrm{M}_\odot/$yr for T~Per.

\textit{NML~Cyg}: We do not observe any circumstellar envelope around NML~Cyg at 31.5 and 37.1~\micron.  Though the DUSTY constant mass-loss models appear to fit the near- to far-IR SED accurately, we cannot conclude from these data alone whether the mass loss around NML~Cyg is smooth and constant, or decreasing over time.  Additionally, at the resolution of FORCAST at 20--40~\micron, we do not observe any evidence for an optically-thick, cool dust shell. The DUSTY models provide an estimate for a mass-loss rate of $\sim4-5\times10^{-4}\,\mathrm{M}_\odot/$yr.  Finally, as described in \cite{schuster2006,schuster2009}, there is an external heat source affecting the temperature structure of the circumstellar envelope surrounding NML~Cyg.  Since DUSTY can only model sources with central internal heating, discrepancies in the mass-loss rates from different measurements probing various parts of the envelope are expected.

In Figure~\ref{fig:MLR}, we summarize the results from this work and Paper~I. We plot the estimated mass-loss rates as a function of luminosity with three mass-loss rate prescriptions from \cite{mauron2011}. \cite{reimers1975} and \cite{kudritzki1978} measured mass-loss rates for O-rich dust-enshrouded RSGs in the LMC and fit an empirical relation to luminosity. Reimers' law is largely consistent with the later formulations, NJ90 \citep{nieuwenhuijzen1990} and observations by \cite{vanloon2005} on dusty RSGs in the LMC at lower luminosities $\left(L\lesssim2\times10^5\,L_\odot\right)$. \cite{mauron2011} apply these mass-loss rate prescriptions to a number of Galactic RSGs. In Figure~\ref{fig:MLR}, we reproduce their implemented formulae at a fixed stellar effective temperature of 3750~K for consistency with their figures.  We adopt their luminosities for all of the sources except IRC~+10420 \citep[$\log{L/L_\odot}\approx5.7-5.8$,][]{debeck2010,tiffany2010}.  The error bars shown in the y-axis (mas-loss rate) are the standard deviation of the derived mass-loss rates from the five best-fitting DUSTY models, and the error bars in the x-axis (luminosity) are from the literature.

We note that the mass-loss rates for the stars in our sample are largely consistent with the Van~Loon and NJ90 prescriptions \citep[analytical forms given in][]{mauron2011}.  As noted in Paper~I, $\mu$~Cep has a curiously low mass-loss rate for its luminosity class, and the hypergiants VY~CMa and IRC~+10420 are well-known, extremely luminous hypergiants whose mass-loss rates are among the highest observed.

DUSTY modeling, in combination with imaging from \SOFIA/FORCAST and MIRAC, are powerful tools for exploring the mass-loss histories and dust density profiles of luminous supergiants.  Although the spatial resolution of FORCAST yielded a PSF too wide for us to compare DUSTY output radial profiles, the photometry from 20--40~\micron\ represents a crucial dataset in constraining the thermal dust properties of RSGs.  The models shown here and in Paper~I hint at possible variable mass-loss rates among the most luminous red supergiants while others have constant mass-loss histories over the past few thousand years. We plan to follow up observations of the supergiants discussed here with high-resolution imaging at $5-12$~\micron\ with LMIRCam and NOMIC on the LBT.  Surface brightness profiles with better spatial resolution at these shorter wavelengths, when coupled with SED modeling, will allow us to characterize mass-loss events from the last few hundred years.

\vspace{4mm}
We thank Rubab Khan for discussion on $\chi^2$-minimization of DUSTY models. This work has used unpublished data from Michael Schuster's PhD thesis, which is available through the SAO/NASA Astrophysics Data System (ADS) at \url{http://adsabs.harvard.edu/abs/2007PhDT........28S}. Financial support for this work was provided by NASA through awards SOF 03 0082 and SOF 04 0013 to R.~M.~Humphreys issued by USRA. RDG acknowledges support from NASA and the United States Air Force.

\facilities{Akari, Herschel (PACS), IRAS, ISO, MMT (MIRAC), Spitzer, SOFIA (FORCAST)}

\software{Astropy \citep{astropy2013}, DUSTY \citep{ivezic1997}, HIPE \citep{ott2010}}

\figEnhanced
\figVXSgrSED
\figVXSgrRad
\figSPerSED
\figSPerRad
\figSPerRadMIRAC
\figRSPerSED
\figRSPerRad
\figTPerSED
\figTPerRad
\figTPerRadMIRAC
\figNMLCygSED
\figNMLCygRad
\figNMLCygRadMIRAC
\figMLR

\end{document}